\documentstyle[prl,preprint,aps,epsf]{revtex}
\begin{document}

\title{Newton Law in DGP Brane-World with Semi-Infinite Extra Dimension}
%\preprint{MCTP-01-19}
\author{ 
D. K. Park$^{1}$\footnote{email: dkpark@hep.kyungnam.ac.kr}, 
S. Tamaryan$^{2}$\footnote{e-mail: sayat@moon.yerphi.am},
Yan-Gang Miao$^{3}$\footnote{e-mail: miao@het.phys.sci.osaka-u.ac.jp},
%H. J. W. M\"{u}ller-Kirsten$^{4}$\footnote{e-mail:
%mueller1@physik.uni-kl.de} 
 }
\address{
1.Department of Physics, Kyungnam University, Masan, 631-701, Korea \\
2.Theory Department, Yerevan Physics Institute, Yerevan-36, 375036, Armenia\\
3. Department of Physics, Osaka University, Toyonaka, Osaka 560-0043, Japan\\
%4. Department of  Physics, University of Kaiserslautern, 
%  67653 Kaiserslautern, Germany
}

\maketitle

%\date{\today}
\maketitle
\begin{abstract}
Newton potential for DGP brane-world scenario is examined when the 
extra dimension is semi-infinite. The final form of the potential involves
a self-adjoint extension parameter $\alpha$, which plays a role of an 
additional mass (or distance) scale. The striking feature of Newton potential
in this setup is that the potential behaves as seven-dimensional in
long range when $\alpha$ is nonzero.
For small $\alpha$ there
is an intermediate range where the potential is five-dimensional.
Five-dimensional Newton constant decreases with increase of $\alpha$ from zero.
In the short range the four-dimensional behavior is recovered. The physical
implication of this result is discussed in the context of the accelerating
behavior of universe. 

\end{abstract}

%---------------------------------------------------------------------------
\newpage
Although the extra dimensional theories have their own long 
history\cite{lee84,ruba83,viss85}, the recent activities on this field seem 
to be motivated from string theories\cite{hora96}. Much cosmological 
implications are investigated and established from the recent brane-world
scenario. Especially Randall-Sundrum(RS) scenario\cite{rs99-1,rs99-2}, one
of the recent brane-world scenario developed by making use of the 
warped extra dimensions, provides an clue for the nature of $1/r$-type
Newton potential in our universe.

The original RS computation of Newton potential is extended and developed 
from various aspects\cite{garr00,gidd00,duff00,park02-1,park02-2}.
Especially, Refs.\cite{park02-1,park02-2} derived Newton potential arising
due to the confined gravity on the brane when bulk space is a single
copy of $AdS_5$ from an aspect of the singular quantum mechanics(SQM).
In this case the gravitational fluctuation equation is treated as an usual 
Schr\"{o}dinger equation with a singular potential and it should be 
solved with incorporation of the self-adjoint extension technique\cite{reed75}.
The real parameter, say $\xi$, introduced in the course of the self-adjoint
extension parametrizes the boundary condition(BC) the gravitational 
fluctuation obeys on the brane.

The SQM approach is ,more recently, applied to the RS scenario when $4d$ induced
gravity is involved\cite{park03-2,park03-3}. 
The physical origin of the $4d$ induced gravity is 
one-loop quantum effect\cite{capp75,adl82,zee82}. When $\xi = 1/2$ which makes
a singular brane to be usual RS brane, the $4d$ induced term generates an
intermediate range in which the $5d$ potential $1/r^2$ emerges. Furthermore, 
the other singular branes corresponding to different $\xi$ may trap a massive
graviton, leading to Yukawa-like gravitational behavior. 

Recently, the brane-world scenario with Minkowski bulk has attracted  
attention, which is often referred as DGP model\cite{dvali00,dvali01}. The 
model also involves a $4d$ induced term and recently applied to the 
cosmological constant hierarchy and the accelerating 
universe\cite{deff01,deff02,dvali03,ruba03,gaba03}.
In this letter we will examine Newton law assigning on the general 
$3$-brane in DGP scenario when the extra dimension is infinite briefly
and semi-infinite in detail. 

For the case of the infinite extra dimension it is well-known that Newton
potential is five-dimensional at long range and four-dimensional at short
range\cite{dvali00}. We will reproduce this result by applying SQM to the 
fluctuation equation. For the case of the semi-infinite extra dimension the 
final form of Newton potential involves a self-adjoint extension
parameter $\alpha$, which makes an additional distance scale
when $\alpha \neq 0$. When $\alpha = 0$, Newton potential is 
similar to that for the case of the infinite extra dimension. However, the 
$5d$ Newton constant $G_5^{\prime}$ becomes $G_5^{\prime} = 2 G_5$ where 
$G_5$ is $5d$ Newton constant derived when the extra dimension is infinite.
The most striking result occurs in the long-range behavior of the potential
when $\alpha > 0$. In this range Newton potential becomes seven-dimensional. 
In the
intermediate range the potential is five-dimensional with smaller Newton
constant than $2 G_5$. The four-dimensional potential is recovered at 
short range. 

Let us start with the Einstein-Hilbert action
\begin{equation}
\label{action1}
S = M^3 \int d^4x dy \sqrt{-G} \tilde{R} + M_p^2 \int d^4x 
\sqrt{-g} R
\end{equation}
where $M$ and $M_p$ are $5d$ and $4d$ Planck scale respectively. 
The curvature scalars $\tilde{R}$ and $R$ are respectively five-dimensional
one and four-dimensional one constructed by $5d$ metric tensor
$G_{MN}(x, y)$ and $4d$ metric tensor 
$g_{\mu \nu}(x) \equiv G_{\mu \nu}(x, y=0)$. Of course, the second term in 
Eq.(\ref{action1}) represents the induced term generated by one-loop 
quantum effect\cite{capp75,adl82,zee82}. 
The equation derived from action (\ref{action1}) is 
\begin{equation}
\label{einstein1}
\left( \tilde{R}_{MN} - \frac{1}{2} G_{MN} \tilde{R} \right)
+ \lambda 
\left( R_{\mu \nu} - \frac{1}{2} g_{\mu \nu} R \right)
\delta_M^{\mu} \delta_N^{\nu} \delta(y) = 0
\end{equation}
where $\lambda = M_p^2 / M^3$. One can easily show that when 
$\lambda = 0$ (or $\infty$), Eq.(\ref{einstein1}) reduces to the usual 
$5d$ (or $4d$) Einstein field equation without the energy-momentum 
tensor. This fact makes us expect to derive the usual $5d$ $ 1/r^2$-type and 
$4d$ $ 1/r$-type gravitational potential when $\lambda = 0$ and 
$\lambda = \infty$ respectively.
The expectation is correct when the extra dimension is infinite.
For the case of the semi-infinite extra dimension, however, 
the solution of the fluctuation
equation is dependent on the BCs, which introduces two different distance
scales. Thus our expectation is not valid in the latter case.

Since there is no contribution from matter, the flat
metric $G_{MN} = \eta_{MN}$ is a trivial solution of Eq.(\ref{einstein1}).
In order to examine the behavior of the fluctuation around the trivial
solution, we define
\begin{equation}
\label{fluc1}
G_{MN} = \eta_{MN} + H_{MN}
\end{equation}
with assumption $|H_{MN}| << 1$. Inserting Eq.(\ref{fluc1}) into 
Eq.(\ref{einstein1}) one can derive the fluctuation equation
\begin{equation}
\label{fluc2}
H_{\mu \nu}^{\prime \prime} + \Box^{(4)} H_{\mu \nu} + 
\lambda \Box^{(4)} H_{\mu \nu} \delta(y) = 0
\end{equation}
where the prime denotes a differentiation with respect to $y$ and 
$\Box^{(4)} \equiv \partial_{\mu} \partial^{\mu}$. When deriving 
Eq.(\ref{fluc2}) we have used the traceless and transverse gauge
\begin{equation}
\label{gauge}
H_{55} = H_{\mu 5} = H_{\mu}^{\mu} = \partial^{\mu} H_{\mu \nu} = 0
\end{equation}
to ignore the tensor structure of $H_{MN}$ for simplicity.
Defining $\psi(y)$ as $\psi(y) \equiv H_{\mu \nu} (x, y) e^{-i p x}$
changes the fluctuation equation (\ref{fluc2}) as a Schr\"{o}dinger-type
equation
\begin{equation}
\label{schro1}
\hat{H} \psi(y) = E \psi(y)
\end{equation}
where 
\begin{equation}
\label{hamil1}
\hat{H} = -\frac{1}{2} \partial_y^2 - \lambda E \delta(y)
\end{equation}
and $E = m^2 / 2 \equiv -p^2 / 2$. 

The remarkable feature of $\hat{H}$ from the aspect of 
SQM is the fact that the coupling constant of the 
singular potential $\delta(y)$ is dependent on the energy eigenvalue.
Similar potential in the fluctuation level arises when the bulk is 
$AdS_5$\cite{park03-2,park03-3}. 
In this case the fixed-energy amplitude can be constructed 
explicitly by applying Schulman procedure\footnote{Schulman procedure is in 
detail explained in Ref.\cite{park02-2}}. 
If, for example, the Hamiltonian
is $H = H_V(\vec{p}, \vec{r}) + \hat{v}(E) \delta(\vec{r})$, the fixed-energy
amplitude $\hat{G}[\vec{r}_1, \vec{r}_2; E]$ for total Hamiltonian can be 
constructed from the fixed-energy amplitude $\hat{G}_V[\vec{r}_1, \vec{r}_2; E]$
for $H_V(\vec{p}, \vec{r})$ as following;
\begin{equation}
\label{schul1}
\hat{G}[\vec{r}_1, \vec{r}_2; E] = \hat{G}_V[\vec{r}_1, \vec{r}_2; E]
- \frac{\hat{G}_V[\vec{r}_1, \vec{0}; E] \hat{G}_V[\vec{0}, \vec{r}_2; E]}
       {\frac{1}{\hat{v}(E)} + \hat{G}_V[\vec{0}, \vec{0}; E]}.
\end{equation}
Once the fixed-energy amplitude for total Hamiltonian system is 
constructed, Newton potential on the brane is directly computed as
following\cite{park03-3};
\begin{equation}
\label{newton}
V(r) = \frac{1}{2 \pi^2 M^3 r}
       \int_0^{\infty} dm m \sin mr \hat{G}[0, 0; \frac{m^2}{2}]
\end{equation}
if the brane is located at $y = 0$. Thus the main problem is converted
to the construction of the fixed-energy amplitude.

If $H_V$ is simply $1d$ free case as Eq.(\ref{hamil1}), the 
construction is considerably
easy if extra dimension $y$ is infinitely flat. If, however, the extra 
dimension is semi-infinite ($y \geq 0$) and the $3$-brane is located at
the end point of it, the construction of $\hat{G}_V$ for $H_V$ is not 
relatively simple. 
In this case $\hat{G}_V$ should be derived by incorporating the self-adjoint
extension. In this letter we will examine Newton potential on the 
brane for both cases.

Firstly, let us discuss the case of the infinite flat extra dimension for 
brevity. In this case the fixed-energy amplitude 
$\hat{G}_V[\vec{r}_1, \vec{r}_2; E]$ in Eq.(\ref{schul1}) is simply 
replaced by an amplitude for usual $1d$ free-particle:
\begin{equation}
\label{free1}
\hat{G}_V[\vec{r}_1, \vec{r}_2; E] \rightarrow
\hat{G}_F[a, b; E] = 
\frac{e^{-\sqrt{2E} |a - b|}}{\sqrt{2E}}.
\end{equation}
Inserting Eq.(\ref{free1}) into Eq.(\ref{schul1}) with letting 
$\hat{v}(E) = -\lambda E$ enables us to compute the fixed-energy
amplitude in this set-up:
\begin{equation}
\label{fixed1}
\hat{G}[a, b; E] = \frac{e^{-\sqrt{2E} |a - b|}}{\sqrt{2E}}
- \frac{1}{\sqrt{2E} - \frac{2}{\lambda}}
e^{-\sqrt{2E} (|a| + |b|)}.
\end{equation}
%Taking an inverse Laplace transform to Eq.(\ref{fixed1}), one can derive a 
%Euclidean Kernel explicitly:
%\begin{equation}
%\label{kernel1}
%G[a, b; t] = G_0[a, b; t] - G_0[|a|, -|b|; t] - 
%\frac{2}{\lambda} \int_0^{\infty} dz e^{\frac{2}{\lambda} z}
%G_0[|a|, -|b| - |z|; t]
%\end{equation}
%where $G_0[a, b; t]$ is an Euclidean Kernel for $1d$ free particle:
%\begin{equation}
%\label{kernel2}
%G_0[a, b; t] = \frac{1}{\sqrt{2 \pi t}}
%e^{-\frac{(x - y)^2}{\sqrt{2E}}}.
%\end{equation}
%It is straightforward to show that $G[a, b; t]$ obeys a boundary condition(BC):
%\begin{equation}
%\label{boundary1}
%\frac{\partial G}{\partial a} [0^+, b; t] - 
%\frac{\partial G}{\partial a} [0^-, b; t] = - 2 \lambda
%\frac{\partial}{\partial t} G[0, b; t].
%\end{equation}
%Thus, the energy-dependence of the coupling constant yields an time-derivative
%in the BC of the Euclidean propagator. This may be related to the principle 
%of the energy-time uncertainty.
To compute Newton potential in this case, let us insert 
\begin{equation}
\label{fixed2}
\hat{G}[0, 0; \frac{m^2}{2}] = 
\frac{-\frac{2}{\lambda}}{m \left(m + \frac{2}{\lambda} \right)}
\end{equation}
into Eq.(\ref{newton}). Then, the potential on the brane reduces to
\begin{equation}
\label{potential1}
V(r) = -\frac{1}{\pi^2 M_p^2 r}
\int_0^{\infty} dm \frac{\sin mr}{m + \frac{2}{\lambda}}.
\end{equation}
When deriving Eq.(\ref{fixed2}) we identified $m$ as $\sqrt{2E} = -m$ to 
derive a correct sign in Newton potential, which seems to correspond to a 
choice of the 
retarded Green's function.

Before computing Newton potential $V(r)$, let us consider two special cases.
If $M=0$, $\lambda$ becomes infinity and $V(r)$ simply reduces to 
four-dimensional:
\begin{equation}
\label{d4potential}
V_{d=4}(r) = - \frac{G_4}{r}
\end{equation}
where
\begin{equation}
\label{d4newton}
G_4 = \frac{1}{2 \pi M_p^2}.
\end{equation}
If $M_p = 0$, $V(r)$ in Eq.(\ref{potential1}) is changed into
\begin{equation}
\label{potential2}
V(r) = -\frac{1}{2 \pi^2 M^3 r} \int_0^{\infty} dm \sin mr.
\end{equation}
Since the integral in Eq.(\ref{potential2}) is not 
well-defined, we should adopt an appropriate regularization. For the
regularization we introduce a damping factor $e^{-\epsilon m}$ as 
following:
\begin{equation}
\label{regu1}
\int_0^{\infty} dm \sin mr \rightarrow
\int_0^{\infty} dm \sin mr e^{-\epsilon m}.
\end{equation}
Then, the final form of Newton potential is purely five-dimensional as 
expected
\begin{equation}
\label{d5potential}
V_{d=5}^{(Reg)}(r) = - \frac{G_5}{r^2}
\end{equation}
where
\begin{equation}
\label{d5newton}
G_5 = \frac{1}{2 \pi^2 M^3}.
\end{equation}
Thus our previous expectation is exactly recovered at the level of Newton 
potential with a correct sign.

The general Newton potential is obtained by carrying out the integral of 
Eq.(\ref{potential1}):
\begin{equation}
\label{potential3}
V(r) = - \frac{1}{\pi^2 M_p^2 r}
\left[ \mbox{ci}\left(\frac{2 r}{\lambda}\right) \sin \left(\frac{2 r}{\lambda}\right) -
\cos \left(\frac{2 r}{\lambda}\right) \mbox{si} \left(\frac{2 r}{\lambda}\right)
                                                     \right]
\end{equation}
where $\mbox{ci} (z)$ and $\mbox{si} (z)$ are usual sine and cosine integral
functions. Using the asymptotic behaviors and the short-range expansions 
of these special functions, it is straightforward to show that the 
long-range behavior($r >> \lambda /2$) of Newton potential is five-dimensional
\begin{equation}
\label{potential4}
V(r) = - \frac{G_5}{r^2}
         \left( 1 - \frac{2 r_0^2}{r^2} \right)
\end{equation}
and the short-range behavior($r << \lambda /2$) is four-dimensional
\begin{equation}
\label{potential5}
V(r) = - \frac{G_4}{r}
\left[ 1 + \frac{2 r}{\pi r_0} 
      \left( \gamma - 1 + \ln \frac{r}{r_0} \right) \right]
\end{equation}
where $r_0 \equiv \lambda / 2$ and $\gamma$ is an Euler's constant.
If, of course, $\lambda = 0$ (or $\infty$), Eq.(\ref{potential4})
(or (\ref{potential5})) exactly coincides with $5d$ (or $4d$) Newton
potential given in Eq.(\ref{d4potential}) and (\ref{d5potential}).

Now, let us discuss Newton potential when the extra dimension is 
semi-infinite. With this set-up the fixed-energy amplitude should be 
computed by incorporating an half-line constraint. From the viewpoint of 
quantum mechanics the constraint should be chosen to maintain the 
unitarity for physical reason. This implies that the BC we should
adopt must yield the vanishing probability current at $y=0$, 
{\it i.e.} 
$(\psi^{\ast} \partial_y \psi - \partial_y \psi^{\ast} \psi)|_0 = 0$.
This requirement can be ensured if all states in the domain of the definition
of the Hamiltonian obey
\begin{equation}
\label{self1}
\frac{\partial}{\partial y} \psi \bigg|_0 = \alpha \psi \bigg|_0
\end{equation}
where $\alpha$ is an arbitrary real number called as a 
`self-adjoint extension parameter'.

The fixed-energy amplitude for free particle living in half-line compatible 
with
BC (\ref{self1}) is computed long ago in Refs.\cite{cla80,far90}, 
whose explicit form
is 
\begin{equation}
\label{fixed3}
\hat{G}_V[\vec{r}_1, \vec{r}_2; E] \rightarrow
\hat{G}_{\alpha}[a, b; E] = \frac{1}{\sqrt{2E}}
\left( e^{-\sqrt{2E} |a - b|} + 
       \frac{\sqrt{2E} - \alpha}{\sqrt{2E} + \alpha}
       e^{-\sqrt{2E} (a + b)}
                                \right).
\end{equation}
Particular attention is paid to $\alpha = 0$ and $\alpha = \infty$ cases:
\begin{eqnarray}
\label{parti1}
\hat{G}_{\alpha=0}[a, b; E]&\equiv& \hat{G}^N[a, b; E]
= \frac{1}{\sqrt{2E}}
\left( e^{-\sqrt{2E} |a - b|} + e^{-\sqrt{2E} (a + b)} \right)
                                                  \\   \nonumber
\hat{G}_{\alpha=\infty}[a, b; E] &\equiv& \hat{G}^D[a, b; E]
= \frac{1}{\sqrt{2E}}
\left( e^{-\sqrt{2E} |a - b|} - e^{-\sqrt{2E} (a + b)} \right).
\end{eqnarray}
One can show easily that $\hat{G}^D$ and $\hat{G}^N$ obey the usual
Dirichlet and Neumann BCs at $y=0$.

Inserting Eq.(\ref{fixed3}) into (\ref{schul1}) with $\hat{v}(E) = -\lambda E$,
it is easy to compute the fixed-energy amplitude for total 
Hamiltonian system. Then, 
the fixed-energy amplitude on the brane simply reduces to 
\begin{equation}
\label{fixed4}
\hat{G}[0, 0; \frac{m^2}{2}] = 
- \frac{2}{\lambda m^2 + m - \alpha}.
\end{equation}
Thus, combining Eq.(\ref{newton}) and Eq.(\ref{fixed4}) one can express the 
gravitational potential $V_{\alpha}(r)$ as following:
\begin{equation}
\label{potential6}
V_{\alpha}(r) = - \frac{1}{\pi^2 M_p^2 (m_+ - m_-) r}
\int_0^{\infty} dm
\left( \frac{m_+}{m + m_+} - \frac{m_-}{m + m_-} \right) \sin mr
\end{equation}
where
\begin{equation}
\label{mpmm}
m_{\pm} = \frac{1}{2 \lambda}
\left[ 1 \pm \sqrt{1 + 4 \lambda \alpha} \right].
\end{equation}
At this stage it is worthwhile noting that our potential $V_{\alpha}(r)$ 
involves two mass (or distance) scales for arbitrary nonzero $\alpha$. 
This means Newton potential for the case of semi-infinite
extra dimension can be completely different from that for the case of the 
infinite extra dimension. Shortly, we will show that the extremely long-range
behavior of the potential in this case is seven-dimensional when
$\alpha > 0$. 

For simplicity let us consider $\alpha = 0$ case first where only one 
distance scale emerges. Since $m_+ = 1/\lambda$ and $m_- = 0$, Newton
potential $V_{\alpha = 0}(r)$ can be obtained straightforwardly from 
Eq.(\ref{potential6}):
\begin{equation}
\label{potential7}
V_{\alpha = 0}(r) = - \frac{1}{\pi^2 M_p^2 r}
\left[ \mbox{ci} \left( \frac{r}{\lambda} \right) 
       \sin \left( \frac{r}{\lambda} \right) - 
       \cos \left( \frac{r}{\lambda} \right)
       \mbox{si} \left( \frac{r}{\lambda} \right) \right].
\end{equation}
Thus, the long-range behavior ($r >> \lambda$) is five-dimensional
\begin{equation}
\label{potential8}
V_{\alpha = 0}(r) = - \frac{2 G_5}{r^2} 
\left( 1 - 2 \frac{\lambda^2}{r^2} \right)
\end{equation}
and the short-range behavior ($r << \lambda$) is four-dimensional
\begin{equation}
\label{potential9}
V_{\alpha = 0}(r) = - \frac{G_4}{r}
\left[ 1 + \frac{2 r}{\pi \lambda}
       \left( \gamma - 1 + \ln \frac{r}{\lambda} \right) \right].
\end{equation}
Thus, the global structure of Newton potential when $\alpha=0$ is very
similar to that in the case of the infinite extra dimension. However, 
$5d$ Newton constant in the long-range behavior becomes 
$G_5^{\prime} = 2 G_5$ while $4d$ Newton constant in the short-range 
behavior remains unchanged. This seems to imply that the change of the 
fifth dimension does not affect the $4d$ quantity.

Next let us consider $\alpha > 0$ case. Since $m_+ > 0$ and $m_- < 0$ in
this case we define two different distance scales 
$r_{\pm} = \pm 1 / m_{\pm}$. The explicit computation of the integrals 
in Eq.(\ref{potential6}) makes Newton potential $V_{\alpha > 0}(r)$ to be
\begin{eqnarray}
\label{potential10}
V_{\alpha > 0}(r)&=& - 
\frac{1}{\pi^2 M_p^2 \left( \frac{1}{r_+} + \frac{1}{r_-} \right) r}
\Bigg[ \frac{1}{r_+}
      \left\{ \mbox{ci} \left( \frac{r}{r_+} \right)
              \sin \left( \frac{r}{r_+} \right)
              - \mbox{si} \left( \frac{r}{r_+} \right)
              \cos \left( \frac{r}{r_+} \right) \right\}
                                                      \\   \nonumber
& & \hspace{1.0cm}
       - \frac{1}{r_-}
         \left\{ \mbox{ci} \left( \frac{r}{r_-} \right)
                 \sin \left( \frac{r}{r_-} \right)
                 - \mbox{si} \left( \frac{r}{r_-} \right)
                  \cos \left( \frac{r}{r_-} \right)
                 - \pi \cos \left( \frac{r}{r_-} \right) \right\}
                                                                  \Bigg].
\end{eqnarray}
Using the asymptotic behaviors of the sine and cosine integral functions
one can show that the long-range behavior ($r >> r_+$, $r >> r_-$) of 
$V_{\alpha > 0}(r)$ is seven-dimensional as following
\begin{equation}
\label{potential11}
V_{\alpha > 0}(r) = - \frac{1}{\pi^2 M^3 \alpha^2 r^4}
\left[ 1 - \frac{12 (r_+^2 + r_-^2)}{r^2} \right].
\end{equation}
%When deriving Eq.(\ref{potential11}), we ignored the oscillating factor 
%of Eq.(\ref{potential10}). 
Mathematically, this pecular behavior arises due to 
the appearance of two different distance scales and the exact cancellation 
of the five-dimensional term. However, the physical origin of this 
seven-dimensional Newton potential is unclear to us. It is interesting to note
that $7d$ Newton constant is contributed from $G_5$ and the self-adjoint
extension parameter as $G_7 = 2 G_5 / \alpha^2$.

Next let us examine the behavior of $V_{\alpha > 0}(r)$ in intermediate
range ($r_+ << r << r_-$). Since $r_- - r_+ = 1 / \alpha$, this region actually
arises only for small $\alpha$. When, thus, $\alpha$ is comparatively large, 
this region does not exist. In this range the potential behaves as
\begin{equation}
\label{potential12}
V_{\alpha > 0}(r) = - \frac{4 G_5}{(1 + \sqrt{1 + 4 \lambda \alpha}) r^2}
\left[ 1 + 
      \left\{ \frac{\pi}{2} \frac{r}{r_-} - 2 \left(\frac{r_+}{r}\right)^2
                                                            \right\}
                                                              \right].
\end{equation}
The leading term of the potential is five-dimensional, whose Newton
constant $G_5^{\prime \prime}$ is less than $2 G_5$:
\begin{equation}
\label{nconst1}
G_5^{\prime \prime} = \frac{4 G_5}
                         {1 + \sqrt{1 + 4 \lambda \alpha}} < 2 G_5.
\end{equation}
Thus $5d$ Newton constant in this region is smaller than that in 
$\alpha = 0$ case. For example, if $\lambda \alpha << 1$, 
$G_5^{\prime \prime}$ becomes
\begin{equation}
\label{nconst2}
G_5^{\prime \prime} \sim 2 G_5
(1 - \lambda \alpha + \cdots).
\end{equation}

The subleading term in this region is not uniquely determined. In the region
$r / r_- >> (r_+ / r)^2$, we have 
\begin{equation}
\label{potential13}
V_{\alpha > 0}(r) = - \frac{G_5^{\prime \prime}}{r^2}
\left( 1 + \frac{\pi}{2} \frac{r}{r_-} \right)
\end{equation} 
and in the region $r / r_- << (r_+ / r)^2$, we have
\begin{equation}
\label{potential14}
V_{\alpha > 0}(r) = - \frac{G_5^{\prime \prime}}{r^2}
\left[ 1 - 2 \left( \frac{r_+}{r} \right)^2 \right].
\end{equation}

In the extremely short-range region ($r << r_+$, $r << r_-$) the potential
behaves as four-dimensional
\begin{equation}
\label{potential15}
V_{\alpha > 0}(r) = - \frac{G_4}{r}
\left[ 1 + \frac{2 r}{\pi}
      \left\{ \frac{\gamma -1}{\lambda} + 
              \frac{\lambda}{\sqrt{1 + 4 \lambda \alpha}}
              \left( \frac{1}{r_+^2} \ln \frac{r}{r_+} - 
                     \frac{1}{r_-^2} \ln \frac{r}{r_-} \right) \right\}
                                                           \right].
\end{equation}
The invariability of $4d$ Newton constant shows again the half-line 
constraint of the extra dimension does not change the $4d$ quantities.

In this letter we examined Newton potential for the DGP brane-world scenario
when the extra dimension is semi-infinite. The final form of Newton potential
involves a self-adjoint extension parameter $\alpha$, which generates an 
additional mass (or distance) scale. The potential for nonzero $\alpha$ 
behaves as seven-dimensional in the extremely long range. In the short range
the potential recovers four-dimensional behavior. If $\alpha$ is very 
small, there is an intermediate range where the potential is five-dimensional.
The $5d$ Newton constant for nonzero $\alpha$ is smaller than that for 
$\alpha =0$ case. It seems to be interesting to understand the physical 
origin of the seven-dimensional behavior in long range.

In Ref.\cite{deff02} the accelerated behavior of universe is approached using
the DGP brane-world picture when the extra dimension is infinite. The origin
of the behavior in this picture comes from weakness of the gravitational 
force in the long range, which is referred as gravity leakage. Since the 
gravitational force becomes weaker for the case of the semi-infinite
extra dimension, the accelerating behavior at ultra large scale may easily
occurred. We hope to address this issue in the future.
\vspace{1cm}

{\bf Acknowledgement}:  
This work was supported by the Kyungnam University
Research Fund, 2003.

\end{document}